\documentclass[prb,preprint,amsmath,amssymb,superscriptaddress]{revtex4}

\usepackage{graphicx}
\usepackage{dcolumn}
\usepackage{bm}

\usepackage{float}

\usepackage{amsmath}

\usepackage{epstopdf}

\begin{document}


\title{ac susceptibility
and electron spin resonance studies of spin dynamics in Ba$_3$NbFe$_3$Si$_2$O$_{14}$:
A geometrically frustrated lattice}

\author{K.-Y. Choi}
\email[]{kchoi@cau.ac.kr}
\affiliation{Department of Physics, Chung-Ang University,  Seoul 156-756, Republic of Korea}

\author{Z.-X. Wang}
\affiliation{Department of Chemistry and Biochemistry, Florida State University, Tallahassee, Florida 32306, USA.}
\affiliation{National High Magnetic Field Laboratory, Florida State University, Tallahassee, Florida 32310, USA}

\author{A. Ozarowski}
\affiliation{National High Magnetic Field Laboratory, Florida State University, Tallahassee, Florida 32310, USA}

\author{J. van Tol}
\affiliation{Department of Chemistry and Biochemistry, Florida State University, Tallahassee, Florida 32306, USA.}
\affiliation{National High Magnetic Field Laboratory, Florida State University, Tallahassee, Florida 32310, USA}

\author{H. D. Zhou}
\affiliation{National High Magnetic Field Laboratory, Florida State University, Tallahassee, Florida 32310, USA}

\author{C. R. Wiebe}
\affiliation{National High Magnetic Field Laboratory, Florida State University, Tallahassee, Florida 32310, USA}
\affiliation{Department of Chemistry, University of Winnipeg, Winnipeg, Manitoba, Canada R3B 2E9}

\author{N. S. Dalal}
\affiliation{Department of Chemistry and Biochemistry, Florida State University, Tallahassee, Florida 32306, USA.}

\bigskip
\bigskip
\bigskip

\begin{abstract}
\bigskip
We report ac susceptibility and high-frequency electron spin resonance (ESR) measurements on the geometrically frustrated compound Ba$_3$NbFe$_3$Si$_2$O$_{14}$ with the N\'{e}el
temperature $T_N=27~K$. An unusually large frequency-dependence of ac susceptibility in the temperature range of 20 - 100~K reveals a spin-glass-like behavior, signalling the presence of
frustration related slow magnetic fluctuations. ESR experiments show a multi-step magnetic and spin chirality ordering process. For temperatures above 30~K, the weak temperature dependence of
the ESR linewidth $\Delta H_{pp}\propto T^{-p}$ with $p=0.8$ evidences the development of short-range correlated spin clusters. The critical broadening with $p =1.8$, persisting down to 14~K,
indicates the coexistence of the short-range ordered spin clusters within a helically ordered state.
Below 9.5~K, the anomalously large decrease of the linewidth reveals the stabilization of a long-range ordered state with one chirality.

\end{abstract}


\maketitle

Geometrically frustrated magnets are currently
attracting intensive research interest in the context of condensed-matter physics.~\cite{FS} Combined effects of quantum fluctuations
and highly degenerate states lead to exotic ground
states, a vortex-induced topological order, a long-range order in a higher-order degree of freedom such as scalar or vector chiral orders, quantum criticality
as well as exotic low-lying excitations.~\cite{Nakatsuji,Helton,Kohno, Bulaevskii,Fujimoto,Wang,Choi,Zhou09,Lumata}

The simplest example of the frustrated antiferromagnets is found
in the two-dimensional (2D) triangular lattice.~\cite{Nakatsuji10}
For the nearest-neighbor antiferromagnetic Heisenberg triangular system,
the ground state is given by  a $120^\circ$ antiferromagnetic order.~\cite{Huse,Bernu,Capriotti}  This invokes the notion of a vector spin chirality, which provides a conceptual framework
for understanding phase transitions on the triangular lattices.

Among the triangular lattices  the iron-based langasite
Ba$_3$NbFe$_3$Si$_2$O$_{14}$ (BNFSO) is the focus of research interest
due to its novel magnetic and multiferroic properties.~\cite{Zhou,Zhou10,Marty,
Marty1,Marty2,Loire,Stock} BNFSO  has a non-centrosymmetric trigonal structure
belonging to the \emph{P321} space group. The Fe$^{3+}$ ions (S=5/2)
form isolated triangular lattices on a hexagonal lattice, which are arranged
to form 2D triangular lattices in the \emph{ab} plane. The neighboring planes are
separated by layers containing Ba and Nb cations.
A magnetic ordering occurs below the N\'{e}el
temperature $T_N=27~K$ despite a large Curie-Weiss temperature of
$\Theta_W=-173~K$, signifying substantial frustration.
The intratriangle interaction of $J_1\approx 9.9$~K is
mediated by superexchange via oxygen anions forming the tetrahedral coordination.
The intertriangle interaction of $J_2\approx 2.8$~K  in the \emph{ab} plane
is mediated by two oxygens (super-superexchange).
Besides, a set of interplane exchange
pathways are needed to stabilize the helical spin structure along the \emph{c} axis
with the period $1/\tau\approx 7$. Even without applying an external field chiral dynamics
is observed from spin-wave excitations
of the helically modulated $120^\circ$ magnetic order.~\cite{Loire}
This means that BNFSO is a rare example
possessing both a triangular chirality and a helical chirality in a single
lattice.

Neutron scattering experiments reveal a multi-step
magnetic ordering process towards a single chirality phase.~\cite{Zhou10}
Vortices ordering is considered to play a crucial
role in relation to the multiferroic properties. Frustrated magnets
show a many-sided face, depending on the time- and spacescales
of employed spectroscopic methods. Thus, it is
important to investigate a spin and chirality ordering process
using a different time window from neutron scattering.
ac susceptibility and electron spin resonance (ESR) are a complimentary
experimental choice because the former can probe a slow
spin dynamics while the latter is sensitive to fast short-range
spin correlations.

In this brief report, we present evidence for the spin-glass-like slow spin dynamics
by ac susceptibility, and for the persistence of short-range
correlated spin chirality states well down to $T_N$ by ESR.
The simultaneous observation of the {\it slow} and the {\it fast}
spin dynamics might be the characteristic feature of frustrated magnets
with a large spin value and spin chirality.

Single crystals of BNFSO were grown using the
travelling-solvent floating-zone technique as described
elsewhere.~\cite{Zhou} The crystals were characterized using
X-ray diffraction, magnetic susceptibility, specific heat, thermal
conductivity, neutron scattering, dielectric and polarization
measurements.~\cite{Zhou,Zhou10} ac susceptibility measurements were performed as a function of temperature using a SQUID magnetometer (Quantum Design MPMS) in a
frequency range $\nu=1- 1000$~Hz in an ac field of 3~Oe and zero dc magnetic field.
The high frequency-ESR spectra were obtained on the ESR spectrometers at the National High Magnetic Field Laboratory.
The temperature dependence was measured using a homodyne/transmission spectrometer at
$\nu=100$~GHz.~\cite{Cage,Hassan}

\begin{figure}[tbp]
\centering
\includegraphics[width=4.5in]{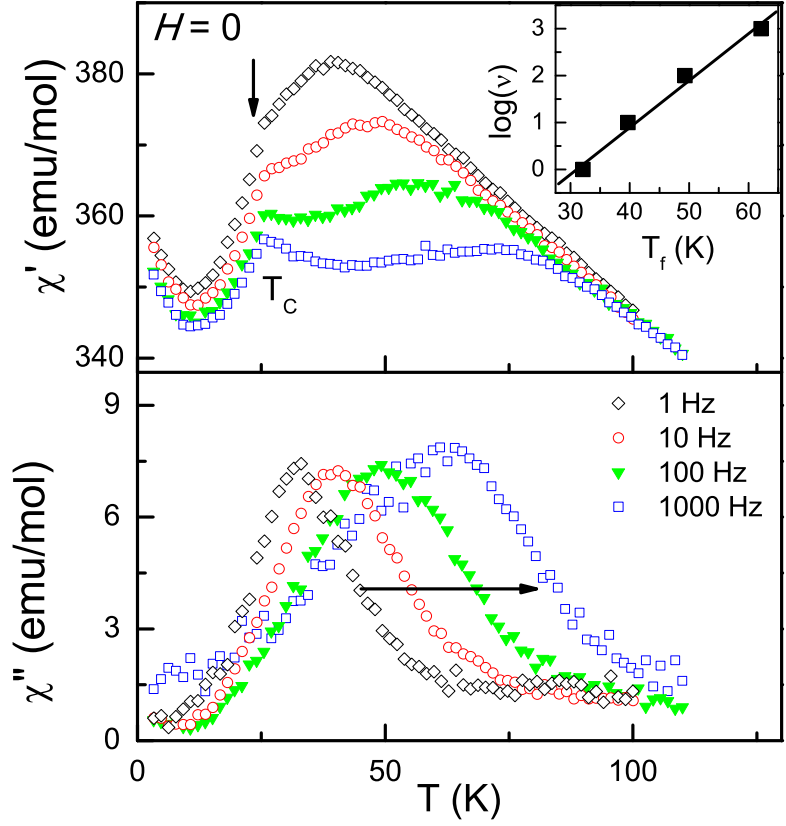}
\caption{Temperature dependence of the real and imaginary parts of the ac susceptibility
at frequencies $\nu=1, 10, 100, \mbox{and}\, 1000$~Hz without external dc field. With increasing frequency the ac susceptibility curve
shifts to higher temperatures (see the arrow). Inset: Frequency
shift of the peak position of the imaginary part of the ac susceptibility versus temperature.} \label{spec1}
\end{figure}

Figure~1 shows the temperature dependence of
the ac susceptibility taken in zero dc magnetic field. With decreasing
temperature the real part of the ac susceptibility, $\chi'(T)$, for $\nu=1$~Hz
exhibits a round maximum at 39~K
and then a kink around $T_N\sim 27$~K. With increasing frequency, the
maximum shifts to higher temperature and decreases in amplitude.
The kink at $T_N$  evolves into a sharp peak without changing the position.
This feature is due to critical fluctuations accompanying a transition to the N\'{e}el ordered state.

The imaginary part of the ac susceptibility, $\chi''(T)$ also exhibits appreciable
frequency dependence between 20 K and 100 K. The loss peak of $\chi''(T)$
is visible for $\nu=1$~Hz at 32~K, which is 7 K lower than that in $\chi'(T)$.
With increasing $\nu$, the peak undergoes a shift to higher temperatures.
and the peak width broadens. In contrast to $\chi'(T)$, there is no decrease in amplitude.
Overall, these features are reminiscent of a spin-glass-like behavior or superparamagnetism.
However, it should be discriminated from conventional spin glasses due to the following reasons.

First, the dc susceptibility does not show any round maximum in contrast to
the ac susceptibility.
Second, the magnitude of the frequency shift is much larger than
what is expected for canonical spin glasses.  We used the well-defined peak position of
$\chi''(T)$ to estimate the shift of the ac susceptibility as a function of
frequency (see the inset of Fig.~1). The curve shifts
by as much as 30 K on varying the frequency from 1 Hz to 1000 Hz. From the maximal
shift $\Delta T_f$ we obtain the value of $\phi=\Delta T/T_f\Delta(\mbox{log}\nu)\sim 0.16$.	 This value is much bigger than that of a spin glass, which is typically the order of $10^{-2}$.~\cite{Mydosh} The large $\phi$ is compatible to some sort of superparamagnetism.

Similar features have been reported in pyrochlores and spin-chain compound
Ca$_3$CoIrO$_6$ with magnetic frustration.
\cite{Matsuhira,Rayaprol}  This suggests that the strong frequency dependence
of the ac susceptibility originates from a slow spin dynamics pertaining to
geometrically frustrated magnets, akin to superparamagnetism.

\begin{figure}[tbp]
\centering
\includegraphics[width=3.2in]{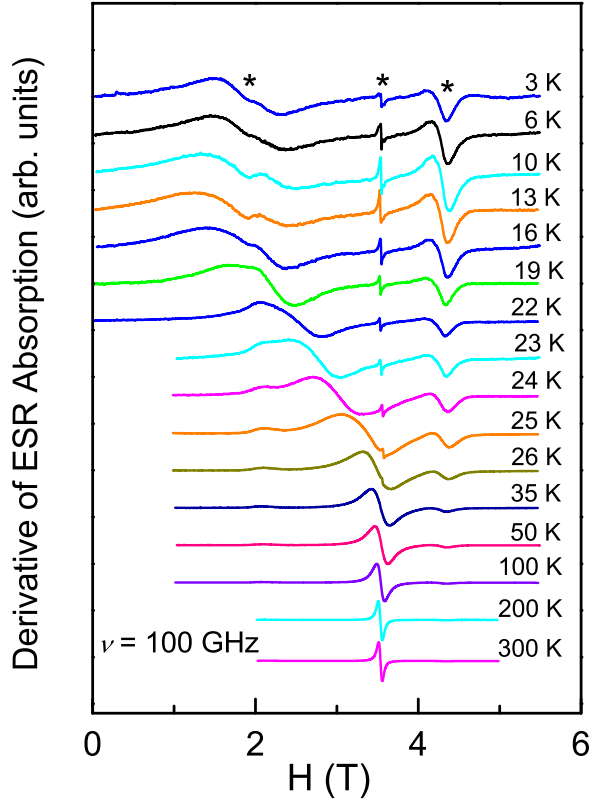}
\caption{Derivative of the ESR absorption of
the BNFSO sample measured at $\nu =100$ GHz as
a function of temperature. The asterisks denote
impurities or parasitic phases, since they exhibit
no shift with temperature.} \label{spec1}
\end{figure}

In addition to the observed slow spin dynamics,
the 2D triangular antiferromagnets are characterized by
fast short-range magnetic correlations over a space scale of
a spin trimer. It is known that high-frequency ESR
can provide direct information on the evolution of spin correlations. Figure~2
shows the temperature dependence of the ESR
spectra measured at 100~GHz. At room temperature,
a narrow single Lorentzian line is observed,
originating from Fe$^{3+}$($3d^5;S=5/2$) ions. The g-factor is close to
the free spin value as expected from a half-filled (3$d^5$) shell.
With decreasing temperature
the spectrum undergoes a broadening and a shift to lower fields.
In addition to the main signal, at low temperatures extra peaks show up.
Their peak positions do not change with temperature and their intensities
show a Curie-like behavior. Thus, they are ascribed to
a few percentage of impurities or parasitic phases. Hereafter, we will focus on the main signal.
To detail the evolution of the spectrum, the resonance field ($H_{res}$)
and the ESR linewidth ($\Delta H_{pp}$) are extracted by fitting the spectra to a derivative of Lorentzian profiles. The results are plotted as  a function of
temperature in Fig.~3.

\begin{figure}[tbp]
\centering
\includegraphics[width=3.9in]{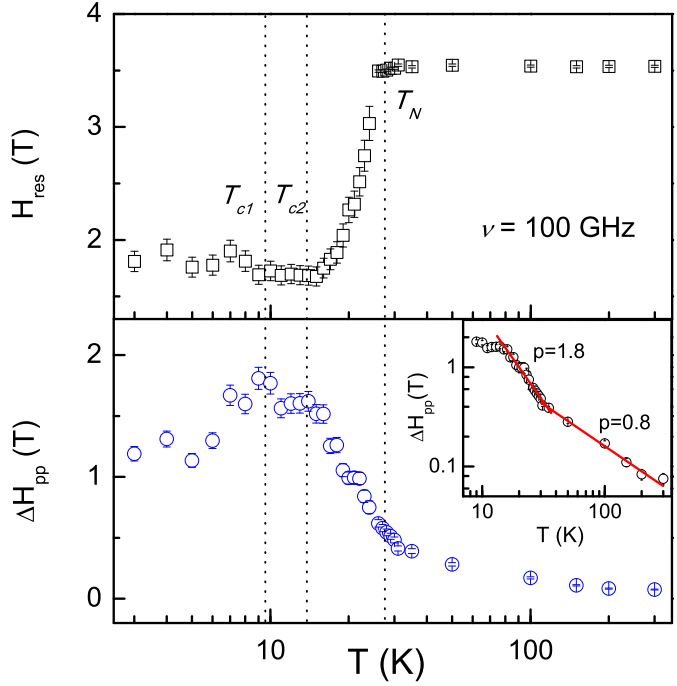}
\caption{Temperature dependences of the resonance field, $H_{res}$,
(upper panel) and the ESR linewidth, $\Delta H_{pp}$ in a semilogarithmic
 plot (lower panel). Inset: A log-log plot of $\Delta H_{pp}$
 versus temperature. The solid lines are a fit to a power law. See
 the text for details. } \label{spec1}
\end{figure}

We observe intriguing features in the temperature dependence.
For temperatures above 30~K ($T>T_N$),  $\Delta H_{pp}$ increases gradually with
lowering temperature while $H_{res}$ hardly varies with temperature.
At temperatures between 30~K and $T_{C2}=14$~K, the linewidth increases
enormously and the resonance field shifts strongly toward lower fields. In the short
temperature interval between $T_{C1}=9.5$~K and $T_{C2}$ both
$\Delta H_{pp}$ and $H_{res}$ show no apparent temperature dependence.
Finally, for temperatures below $T_{C1}$, $H_{res}$ does not change
with temperature while $\Delta H_{pp}$ drops but it remains at a large value
even at the lowest temperature.

To analyze quantitatively the temperature dependence of the linewidth,
$\Delta H_{pp}$ is plotted in a log-log scale as shown in the inset
of Fig.~3. We find the weak power law dependence, $\Delta H_{pp}\propto
T^{-p}$ with $p=0.8$ for $T>30$~K and the strong temperature dependence
with $p=1.8$ between $T_{C2}$ and $T_{N}$.

For conventional magnets, the linewidth is temperature
independent in a high-temperature paramagnetic regime
and starts to broaden in the vicinity of magnetic ordering temperature.
In contrast, frustrated magnets exhibit significant
temperature dependence of the linewidth in a wide temperature range
reaching up to a dozen times the ordering temperature. It is
due to short-range spin correlations over a spin cluster unit.
In our case, the neutron experiments demonstrated the persistence of correlated
magnetic scattering with a vector chiral component up to 100~K~\cite{Stock}
and the presence of diffuse scattering at an order of the Curie-Weiss temperature,
$\Theta_W$.~\cite{Zhou10} Noticeably, the 2D triangular-lattice NiGa$_2$S$_4$ with a low spin
value S=1 showed pronounced T-dependence of $\Delta H_{pp}$ with the critical
exponent $p=2.5$.~\cite{Yamaguchi} The observed smaller exponent for BNFSO suggests that the spin dynamics and critical fluctuations of the studied compound are in proximity to the classical
limit due to a larger spin value $S=5/2$. This might be responsible for the observation
of the slow spin dynamics by ac susceptibility.

An unconventional feature is seen as temperature approaches $T_N$.
The large shift of $H_{res}$ to a lower field for $T<T_N$
implies the development of an internal magnetic field.
This is what is expected for a magnetically ordered state. Instead of the expected line
narrowing, however, the linewidth shows a critical broadening $\Delta H_{pp}\propto
T^{-p}$ with $p=1.8$, which starts at about $30$~K, higher than $T_N$ and persists down
to 14~K.  The onset temperature reminds us of the temperature scale of
40~K, at which the neutron measurements show a development of long-range correlations
and the heat capacity and thermal-conductivity measurements show broad anomalies.~\cite{Zhou,Zhou10} In addition, the neutron scattering study revealed the presence of the diffusive scattering
down to 20~K, which is close to $T_{C2}$. This feature is ascribed to the coexistence of short-range ordered spin clusters within the helically ordered state as has been reported for other frustrated magnets.~\cite{Mirebeau} In our case, it might be due to spin clusters of opposite chirality generated by defects and impurities.

In a magnetically ordered state the ESR linewidth
is determined by four magnon scattering processes and an occupation number of magnon excitations. Thus, with lowering temperature the linewidth decreases with a power law.
BNFSO shows the drop of $\Delta H_{pp}$ at about $T_{C1}\approx 9.5$~K.  This suggests that
a true long-range ordered state with one chirality sets in at $T_{C1}$. However, the large linewidth at the lowest temperature of 3~K indicates that there are the significant fluctuations of the
uniform spin chirality in the measured time scale of  $\nu=100$~GHz.
Finally, we turn to the transit temperature interval between  $T_{C1}$ and $T_{C2}$.
This rather flat feature in $\Delta H_{pp}$ seems to indicate the competition between the narrowing and the broadening mechanism. Unlike the S=1 triangular compound NiGa$_2$S$_4$,
we find no hint for the bound state of a Z$_2$ vortex.

In summary, we have presented a combined ac susceptibility and ESR study of Ba$_3$NbFe$_3$Si$_2$O$_{14}$. We observe a spin-glass-like behavior of the ac susceptibility
in a wide temperature range of 20 - 100~K,  but with much larger
Mydosh parameter, suggesting superparamagnet due to locally ordered clusters.
The ESR measurements reveal the persistence
of a short-range correlated state to a long-range ordered one. The concomitant occurrence of
frustration related {\it slow} and {\it fast} spin dynamics might be a generic
feature of frustrated magnets with a large spin value and a well-defined spin chirality.
For more quantitative understanding, we need systematic studies
of the spin dynamics in the $10^3 - 10^{11} \mbox{s}^{-1}$.

National High Magnetic Field Laboratory is supported by NSF Cooperative Agreement No. DMR-0654118, and by the State of Florida.  K.Y.C. acknowledges financial support from the Priority Research Center Program funded by Korea NRF Grant No. 2009-0093817. C.W. acknowledges the funding through NSERC and the ACS Petroleum Fund.

\end{document}